%Paper: hep-ph/9312203
%From: meissner@crnvax.in2p3.fr
%Date: Wed, 1 Dec 1993 12:07:35 +0100

\magnification = 1200
\def\lapp{\hbox{$ {     \lower.40ex\hbox{$<$}
                   \atop \raise.20ex\hbox{$\sim$}
                   }     $}  }
\def\rapp{\hbox{$ {     \lower.40ex\hbox{$>$}
                   \atop \raise.20ex\hbox{$\sim$}
                   }     $}  }
\def\bar#1{{\not\mathrel #1}}

\def\stackrel#1{\vbox{\ialign{\hfil##\hfil\crcr
           $\raise0.3pt\hbox{$\scriptstyle \leftrightarrow$}$\crcr\noalign
           {\kern-0.02pt\nointerlineskip}
%          {\kern-0.06pt\nointerlineskip}
$\displaystyle{#1}$\crcr}}}
\def\upar#1{\vbox{\ialign{\hfil##\hfil\crcr
           $\raise0.3pt\hbox{$\scriptstyle \leftrightarrow$}$\crcr\noalign
           {\kern-0.02pt\nointerlineskip}
$\displaystyle{#1}$\crcr}}}
\def\ular#1{\vbox{\ialign{\hfil##\hfil\crcr
           $\raise0.3pt\hbox{$\scriptstyle \leftarrow$}$\crcr\noalign
           {\kern-0.02pt\nointerlineskip}
$\displaystyle{#1}$\crcr}}}

\def\g5{\gamma_5}

\def\vev{<\overline{q}q>}

\def\el{{\cal L}}
\def\pa{$\pi {-}a_1$}
\def\mc{\widehat{m}}
\def\ref#1{\relax}
\def\cite#1{\relax}
\topskip=0.60truein
\leftskip=0.18truein
\vsize=8.0truein
\hsize=6.0truein
\tolerance 10000
\hfuzz=20pt

\baselineskip 12pt plus 1pt minus 1pt
\pageno=0

\centerline{\bf THE MOMENTUM--SPACE BOSONIZATION}
\centerline{\bf OF THE NAMBU--JONA-LASINIO MODEL}
\centerline{\bf WITH VECTOR AND AXIAL-VECTOR MESONS}
\vskip 2.5em
\centerline{V\'eronique Bernard, Ulf-G. Mei{\ss}ner}
\vskip .5em
\centerline{\it
Centre de Recherches Nucl\'{e}aires
et Universit\'{e}
Louis Pasteur de Strasbourg}
\centerline{\it
Physique Th\'{e}orique}
\centerline{\it
BP 20Cr,
67037 Strasbourg Cedex 2, France}
\vskip 0.8em
\centerline{and}
\vskip 0.8em
\centerline{A.A. Osipov}
\vskip .5em
\centerline{\it Joint Institute for Nuclear Research,
                Laboratory of Nuclear Problems}
\centerline{\it 141980 Dubna, Moscow Region, Russia}
\vskip 6em
\baselineskip 12pt plus 1pt minus 1pt
\centerline{ABSTRACT}
\medskip
The momentum-space bosonization method is extended to the case
of a Nambu--Jona-Lasinio type model with vector and axial-vector
mesons. The method presented gives the possibility of deriving any meson vertex
function to all orders in  momenta and to the leading order in $1/N_c$.
Two-point functions, which describe one-particle transitions
to the hadronic vacuum, and meson self-energies are considered.
We find new relations which generalize the well-known KSFR relation
and both the first and the second Weinberg sum rules. These
result from a consistent treatment of higher order terms in the momentum
expansion.

\vfill
\noindent CRN 93--57 \hfill  December 1993
\eject

\baselineskip 12pt plus 1pt minus 1 pt
\noindent{\bf 1. Introduction}
\medskip
The Nambu--Jona-Lasinio (NJL) model in a version incorporating all
essential QCD symmetries may be a reasonable low-energy approximation
for QCD. In our previous paper [1] we suggested a systematic method
for evaluating any mesonic N-point function in the bosonized NJL model.
The main idea consists in the
construction of special bosonic variables
to be used for the
description of the observable mesonic states. As a result,
it is possible to extend the usual treatment of bosonized NJL
models, which was formulated in papers [2] and developed in
Refs.~[3]-[6]. The standard approach is essentially oriented to the
derivative expansion of the effective meson Lagrangian.  Our method does
not use this approximation. Therefore, one can expect it to be
a powerful tool for exploration of the extended NJL model, which
includes not only scalar and pseudoscalar fields but also heavier
vector and axial-vector meson states.

One of the principal difficulties we come across on going over to higher
energies that are typical of vector particles is the confinement
property of QCD. The NJL model enables one to see dynamic symmetry
breaking mechanism at work, which leads to meson formation from
quark-antiquark pairs, but it does not forbid emission of
constituent quarks (with mass $m$) into the
continuum. The model has to deal
with this phenomenon already at energies $p^2\sim 4m^2\sim m^2_{\rho}$.
In its formal part, our method does not require an a priori
solution to the
confinement problem. It only covers the general bosonization scheme
for the model and mainly applies to the separation procedure of the
collective
degrees of freedom in theories with four-fermion interaction. For that
aspect the confinement mechanism is not of relevance.
Our goal here is to construct a formal scheme which
involves the bosonization procedure and has all the advantages of the pure
fermionic approach (Hartree-Fock plus Bethe-Salpeter approximation).
The advantages of this approach are the
explicit use of boson variables for
describing the
dynamics of collective excitations and the possibility of gaining
full information on the momentum dependence of vertex functions.

The NJL model belongs to the set of nonrenormalizable theories.
Hence, to
define it completely as an effective model, a regularization scheme must
be specified to deal with the quark-loop integrals in harmony with general
symmetry requirements. As a result, an  additional parameter $\Lambda$
appears, which characterizes the scale of the quark-antiquark forces
responsible for the dynamic chiral symmetry breaking. From the meson mass
spectrum it is known that $\Lambda\sim 1\,\hbox{GeV}$. Here, we will make
use of the
Pauli--Villars [7] regularization, which preserves gauge invariance.
In this form      it was used in  Refs.[8].

For simplicity we consider the linear bosonized version of the extended
NJL model with $U(2)\otimes U(2)$ symmetry which can be explicitly violated
by the current quark masses. The extension to the case of $U(3)\otimes
U(3)$ symmetry will be done elsewhere.

\medskip
\noindent{\bf 2. Momentum-space bosonization of the extended NJL model}
\medskip
Consider the extended $U(2)\otimes U(2)$ NJL Lagrangian with a
local four-quark interaction
$$
\eqalignno{
\el (q)
       &=\overline{q}(i\bar{\partial}-\widehat{m})q
         +{G_S\over 2}
         \left[(\overline{q}\tau_aq)^2+
         (\overline{q}i\gamma_{5}\tau_aq)^2
         \right]\cr
       &\qquad -{G_V\over 2}
        \left[(\overline{q}\gamma^{\mu}\tau_aq)^2+
              (\overline{q}\gamma_{5}\gamma^{\mu}\tau_aq)^2
        \right],\quad&(2.1)\cr
}
$$
where $\overline{q}=(\overline{u}, \overline{d})$
are coloured current quark fields with current mass
$\widehat{m}={\rm diag}(\widehat{m}_u, \widehat{m}_d)$, $\tau_a=
(\tau_0, \tau_i),\, \tau_0=I,\, \tau_i\, (i=1, 2, 3)$ are the Pauli
matrices of the flavour group $SU(2)_f$. The constants of the four-quark
interactions are $G_S$ for the scalar and pseudoscalar cases,
$G_V$ for the vector and the axial-vector cases.
The current mass term explicitly violates the $U(2)\otimes U(2)$ chiral
symmetry of the Lagrangian (2.1). In what follows, we shall only consider
the isospin symmetric case $\widehat{m}_u=\widehat{m}_d=\widehat{m}$.
Introducing boson fields
in the standard way, the Lagrangian takes the form
$$
\eqalignno{
\el (q, \overline{\sigma}, \tilde{\pi}, \tilde{v}, \tilde{a})
    &=\overline{q}
      \left(i\bar{\partial}-\widehat{m}
            +\overline{\sigma}
            +i\g5\tilde{\pi}
            +\gamma^{\mu}\tilde{v}_{\mu}
            +\g5\gamma^{\mu}\tilde{a}_{\mu}
      \right)q\cr
    & -{\overline{\sigma}^2_{a}+\tilde{\pi}^2_{a}\over 2G_S}
      +{\tilde{v}^2_{\mu a}+\tilde{a}^2_{\mu a}\over 2G_V}.
      \quad&(2.2)\cr
}
$$
Here $\overline{\sigma}=\overline{\sigma}_a\tau_a,\quad
\tilde{\pi}=\tilde{\pi}_a\tau_a,
\tilde{v}_{\mu}=\tilde{v}_{\mu a}\tau_a,
\tilde{a}_{\mu}=\tilde{a}_{\mu a}\tau_a$.
The vacuum expectation value of the scalar field
$\overline{\sigma}_0$ turns out to be different from zero
$(<\overline{\sigma}_0> \bar{=}0)$. To obtain the physical field
$\tilde{\sigma}_0$ with $<\tilde{\sigma}_0>=0$ one performs a field shift
leading to a new quark mass $m$ to be identified with the mass of the
constituent quarks
$$
\overline{\sigma}_0-\widehat{m}=\tilde{\sigma}_0-m,\qquad
\overline{\sigma}_i=\tilde{\sigma}_i,
\eqno(2.3)
$$
where $m$ is determined from the gap equation (see Eq.(2.6) below).

Let us integrate out the quark fields
in the generating functional associated with the
Lagrangian (2.2). Evaluating the resulting
quark determinant by a loop expansion one obtains
$$
\eqalignno{
\el (\tilde{\sigma}, \tilde{\pi}, \tilde{v}, \tilde{a})
    &=-i{\rm Tr}\ln
      \left[1+
      (i\bar\partial -m)^{-1}
      (\tilde{\sigma}+i\g5\tilde{\pi}+\gamma^{\mu}\tilde{v}_{\mu}
      +\g5\gamma^{\mu}\tilde{a}_{\mu})
      \right]_{\Lambda}\cr
    & -{\overline{\sigma}^2_{a}+\tilde{\pi}^2_{a}\over 2G_S}
      +{\tilde{v}^2_{\mu a}+\tilde{a}^2_{\mu a}\over 2G_V}.
      \quad&(2.4)\cr
}
$$
The index $\Lambda$ indicates that a regularization of the divergent
loop integrals is introduced. We apply here the Pauli--Villars
regularization [7], which preserves vector gauge invariance and
at the same time might allow to reproduce
the quark condensate for physical values of the current
quark mass [6].
The Pauli--Villars cut-off $\Lambda$ is introduced
by the following replacements
$$
\eqalign{
         e^{-im^2z}
         &\rightarrow R(z)=e^{-im^2z}\left[1-(1+iz\Lambda^2)
          e^{-iz\Lambda^2}\right],\cr
          m^2e^{-im^2z}
         &\rightarrow iR'(z)
          =m^2R(z)-iz\Lambda^4e^{-iz(\Lambda^2+m^2)},\cr}
\eqno(2.5)
$$
where the minimal number of Pauli--Villars regulator
has been introduced.
In this case the expressions for some loop integrals $I_i$ (see
formulae (2.7), (2.11), and (2.24)) coincide with those obtained by the
usual covariant cut-off scheme.

Consider the first terms of the expansion (2.4). From the requirement
for the terms linear in $\tilde{\sigma}$ to vanish we get a modified gap
equation
$$
m-\widehat{m}=8mG_PI_1.
\eqno(2.6)
$$
The integral $I_1$ is equal to
$$
I_1=iN_c\int^{\Lambda}{d^4q\over (2\pi )^4
    (q^2-m^2)}={3\over (4\pi )^2}\left[\Lambda^2-
    m^2\ln\left(1+{\Lambda^2\over m^2}\right)\right],
\eqno(2.7)
$$
where
$N_c=3$ is
the number of colours.

The terms quadratic in the boson fields lead to the amplitudes
$$
\eqalignno{
\Pi ^{\tilde{\pi}\tilde{\pi}}(p^2)&=
    \left[8I_1-G^{-1}_S+p^2g^{-2}(p^2)\right]
    \varphi^+_{\tilde{\pi}^a}\varphi^-_{\tilde{\pi}^a},&(2.8a)\cr
\Pi ^{\tilde{\sigma}\tilde{\sigma}}(p^2)&=
    \left[8I_1-G^{-1}_S+(p^2-4m^2)g^{-2}(p^2)\right]
    \varphi^+_{\tilde{\sigma}^a}\varphi^-_{\tilde{\sigma}^a},&(2.8b)\cr
\Pi ^{\tilde{v}\tilde{v}}(p^2)&=
    \left[g^{\mu\nu}G_V^{-1}+4(p^{\mu}p^{\nu}-g^{\mu\nu}p^2)
    g_V^{-2}(p^2)\right]\varepsilon^{*\tilde{v}^a}_{\mu}(p)
    \varepsilon^{\tilde{v}^a}_{\nu}(p),&(2.8c)\cr
\Pi ^{\tilde{a}\tilde{a}}(p^2)
    &=\left[ g^{\mu\nu}\left( G_V^{-1}+4m^2g^{-2}(p^2)\right)
      \right.\cr
    &\quad +\left. 4(p^{\mu}p^{\nu}-g^{\mu\nu}p^2)
    g_V^{-2}(p^2)\right]\varepsilon^{*\tilde{a}^a}_{\mu}(p)
    \varepsilon^{\tilde{a}^a}_{\nu}(p),\quad&(2.8d)\cr
\Pi ^{\tilde{\pi}\tilde{a}}(p^2)&=
    2img^{-2}(p^2)p^{\mu}\varepsilon^{*\tilde{a}^a}_{\mu}(p)
    \varphi^-_{\tilde{\pi}^a},&(2.8e)\cr
\Pi ^{\tilde{a}\tilde{\pi}}(p^2)&=-
    2img^{-2}(p^2)p^{\mu}\varepsilon^{\tilde{a}^a}_{\mu}(p)
    \varphi^+_{\tilde{\pi}^a}.&(2.8f)\cr}
$$
Here $\varepsilon^{\tilde{v}^a}_{\mu}(p), \varepsilon^{\tilde{a}^a}_{\mu}(p)$
are the polarization vectors of the vector and axial-vector fields.
We have introduced the symbols
$\varphi^-_{\tilde{\pi}^a}=1\quad\hbox{and}\quad
\varphi^-_{\tilde{\sigma}^a}=1$ to explicitely show
the pseudoscalar and
scalar field contents of the pertinent two-point functions.
The functions $g(p^2)$ and $g_V(p^2)$ are determined
by the following integrals
$$
g^{-2}(p^2)= 4I_2 (p^2),
\eqno(2.9)
$$
$$
g^{-2}_V(p^2)={2\over 3}J_2 (p^2),
\eqno(2.10)
$$
$$
\eqalignno{
           I_2(p^2)
           &={3\over 16\pi^2}\int\limits_0^1dy
            \int\limits_0^{\infty}{dz\over z}R(z)
            e^{iz{p^2\over 4}(1-y^2)},&(2.11)\cr
           J_2(p^2)
           &={9\over 32\pi^2}\int\limits_0^1dy(1-y^2)
            \int\limits_0^{\infty}{dz\over z}R(z)
            e^{iz{p^2\over 4}(1-y^2)}.&(2.12)\cr}
$$

Let us diagonalize the quadratic form $(2.8a)+(2.8d)+(2.8e)+(2.8f)$
redefining the axial fields
$$
\eqalign{
         \varepsilon^{\tilde{a}^a}_{\mu}(p)
          &\rightarrow
           \varepsilon^{\tilde{a}^a}_{\mu}(p)-i\beta (p^2)p_{\mu}
           \varphi^-_{\tilde{\pi}^a},\cr
         \varepsilon^{*\tilde{a}^a}_{\mu}(p)
          &\rightarrow
           \varepsilon^{*\tilde{a}^a}_{\mu}(p)+i\beta (p^2)p_{\mu}
           \varphi^+_{\tilde{\pi}^a}.\cr}
\eqno(2.13)
$$
This
determines the function $\beta (p^2)$,
$$
\beta (p^2)={8mI_2(p^2)\over G_V^{-1}+16m^2I_2(p^2)}.
\eqno(2.14)
$$
Consequently, one has no more mixing
between pseudoscalar and axial-vector fields.
The self-energy of the pseudoscalar field takes the form
$$
\Pi ^{\tilde{\pi}\tilde{\pi}}(p^2)=
    \left[8I_1-G^{-1}_S+p^2g^{-2}(p^2)
    \left(1-2m\beta(p^2)\right)\right]
    \varphi^+_{\tilde{\pi}^a}\varphi^-_{\tilde{\pi}^a}.
\eqno(2.15)
$$

Now we can construct special boson variables that will describe
the observed mesons. These field functions ($\pi _a, \sigma _a, v_a, a_a$)
correspond to bound quark-antiquark states and are derived via the
following transformations
$$\eqalign{
\tilde{\pi}^a(p)&=Z^{-1/2}_{\pi}g_{\pi}(p^2)\pi^a(p),\cr
\tilde{\sigma}^a(p)&=Z^{-1/2}_{\sigma}g(p^2)\sigma^a(p),\cr
\tilde{v}^a(p)&={1\over 2}Z^{-1/2}_{v}g_V(p^2)v^a(p),\cr
\tilde{a}^a(p)&={1\over 2}Z^{-1/2}_{a}g_V(p^2)a^a(p),\cr}
\eqno(2.16)
$$
where
$$
g_{\pi}(p^2)={g(p^2)\over \sqrt{1-2m\beta (p^2)}}
            =g(p^2)\sqrt{1+16m^2G_VI_2(p^2)}.
\eqno(2.17)
$$
The new bosonic fields have the self-energies
$$
\eqalign{
\Pi ^{\pi , \sigma}_{ab}(p^2)&=\delta _{ab}Z^{-1}_{\pi , \sigma}
    \left[p^2-m^2_{\pi , \sigma}(p^2)\right],\cr
\Pi ^{v, a}_{\mu\nu , ab}(p^2)&=\delta _{ab}Z^{-1}_{v, a}
    \left\{p_{\mu}p_{\nu}-g_{\mu\nu}\left[ p^2-m^2_{v, a}(p^2)\right]
    \right\}.\cr}
\eqno(2.18)
$$
The $p^2$-dependent masses are equal to
$$
\eqalignno{
m^2_{\pi} (p^2)&=(G^{-1}_S-8I_1)g^2_{\pi}(p^2),&(2.19a)\cr
%               =\hat{m}(mG)^{-1}g^2(p^2),
m^2_{\sigma} (p^2)
               &=\left[1-2m\beta (p^2)\right]m^2_{\pi}(p^2)+4m^2,&(2.19b)\cr
m^2_v(p^2)&={g^2_V(p^2)\over 4G_V}={3\over 8G_VJ_2(p^2)},&(2.19c)\cr
m^2_a(p^2)&=m^2_v(p^2)+6m^2{I_2(p^2)\over J_2(p^2)}.&(2.19d)\cr}
$$
The constants $Z_{\pi , \sigma , v, a}$
are determined by the requirement that the inverse meson field
propagators $\Pi^{\pi , \sigma , v, a}(p^2)$ satisfy the
normalization conditions
$$
\eqalign{
\Pi ^{\pi , \sigma}(p^2)&=
     p^2-m^2_{\pi , \sigma}+{\cal O}\left(
    (p^2-m^2_{\pi , \sigma})^2\right),\cr
\Pi ^{v, a}_{\mu\nu}(p^2)&=-g_{\mu\nu}\left[
     p^2-m^2_{v, a}+{\cal O}\left(
    (p^2-m^2_{v, a})^2\right)\right],\cr}
\eqno(2.20)
$$
around the physical mass points $p^2=m^2_{\pi ,\sigma , v, a}$,
respectively. The conditions (2.20) lead to the values
$$
\eqalignno{
Z_{\pi}&=1+{m^2_{\pi}[1-2m\beta (m^2_\pi )]\over I_2(m^2_{\pi})}
   {\partial I_2(p^2)\over \partial p^2}
   \bigg\vert_{p^2=m^2_{\pi}},&(2.21a)\cr
Z_{\sigma}&=1+{
   m^2_{\sigma}-4m^2\over I_2(m^2_{\sigma})}
   {\partial I_2(p^2)\over \partial p^2} \bigg\vert_{p^2=
   m^2_{\sigma}},&(2.21b)\cr
Z_v&=1+{m^2_v\over J_2(m^2_v)}
   {\partial J_2(p^2)\over \partial p^2} \bigg\vert_{p^2=
   m^2_v},&(2.21c)\cr
Z_a&=1+{m^2_a\over J_2(m^2_a)}
   {\partial J_2(p^2)\over \partial p^2} \bigg\vert_{p^2=
   m^2_a}-{6m^2\over J_2(m^2_a)}
   {\partial I_2(p^2)\over \partial p^2} \bigg\vert_{p^2=
   m^2_a}.&(2.21d)\cr}
$$

Using the expressions (2.18), one can obtain the two-point Green
functions $\Delta (p)$. For example, in the scalar and vector field case
the relations
$$
\eqalign{
&\Pi^{\sigma}_{ab}(p^2)\Delta^{\sigma}_{bc}(p^2)=\delta_{ac},\cr
&\Pi^v_{\mu\nu ,ab}(p^2)\Delta^{v, \nu\sigma}_{bc}(p)=\delta_{ac}
      \delta^\sigma_\mu .\cr}
\eqno(2.22)
$$
give
$$
\eqalignno{
\Delta^{\sigma}_{ab}(p^2)&={\delta_{ab}Z_\sigma\over p^2-m^2_\sigma (p^2)},
&(2.23a)\cr
\Delta^{v, \mu\nu}_{ab}(p)&={\delta_{ab}Z_v\over m^2_v(p^2)}
{p^\mu p^\nu -g^{\mu\nu}m^2_v(p^2)\over p^2-m^2_v(p^2)}.
&(2.23b)\cr}
$$
The formal scheme developed here gives the possibility of evaluating
any mesonic N-point Green function through the parameters of the model:
$\Lambda , m, G_S, G_V,$ and the one-loop integrals $I_1,\, I_2,\, J_2,\,
I_3,\,\ldots\, I_N$. Two examples of these are
$$
\eqalign{
I_2(p^2)&=\int^{\Lambda}{d^4q\over (2\pi )^4}
          {(-iN_c)\over (q^2-m^2)[(q+p)^2-m^2]},\cr
I_3(p_1, p_2)&=\int^{\Lambda}{d^4q\over (2\pi )^4}
          {(-iN_c)\over (q^2-m^2)[(q+p_1)^2-m^2]
          [(q+p_2)^2-m^2]}.\cr}
\eqno(2.24)
$$
This picture corresponds to the       calculations in the framework of the
pure fermionic NJL model where the Bethe-Salpeter equation
sums an infinite class of fermion bubble diagrams.

\medskip
\noindent{\bf 3. Scale-invariant relations and matching conditions}
\medskip
The purpose of this section is to consider some general consequences
of our approach and to compare them with the well-known current algebra
results. We start with the form factor $f_\pi (p^2)$, which describes
the week pion decay $\pi\rightarrow l\nu_l$ and can be expressed in the
following form
$$
\eqalign{
f_\pi (p^2)={m\over \sqrt{Z_\pi}g_\pi (p^2)}.\cr}
\eqno(3.1)
$$
{}From here on, when omitting an argument of a running coupling constant
or a running mass,
we always assume that its value is taken on the mass-shell of the
corresponding particle. The symbol of this particle will be used for that.
For example, on the pion mass-shell $f_\pi (m^2_\pi )=f_\pi =93.3$ MeV.
Combining Eq.(2.19a) for the pion mass and the
gap equation (2.6) with (3.1)
one finds
$$
\eqalign{
m^2_\pi (p^2)f^2_\pi (p^2)={\mc m\over Z_\pi G_S}=-{2\mc\vev\over
     Z_\pi\left(1-{\mc\over m}\right)}.\cr}
\eqno(3.2)
$$
The right-hand side of this equality does not depend on $p^2$. This is
an example of a scale-invariant relation that can be found in the model
under consideration. It extends the well-known current-algebra result
$m^2_\pi f^2_\pi =-2\mc\vev$ derived by Gell-Mann, Oakes and Renner [9]
that is exact at the lowest order of chiral expansion (in powers of external
momenta $p^2$ and quark masses).

Another relation can be found in the sector of vector mesons. The form
factor of the electromagnetic $\rho\rightarrow\gamma$ transition is
equal to
$$
\eqalign{
{1\over f_\rho (p^2)}={2g_\rho (p^2)\over 3\sqrt{Z_\rho}}J_2(p^2).\cr}
\eqno(3.3)
$$
Using this formula and Eq.(2.19c) one can obtain the following
scale-invariant result
$$
\eqalign{
{m^2_\rho (p^2)\over f^2_\rho (p^2)}={1\over 4Z_\rho G_V}.\cr}
\eqno(3.4)
$$
It is interesting to note here that the constant $G_V$ is related to
the pion decay form factor too. This is a direct consequence of the \pa\
mixing,
$$
\eqalign{
{f^2_\pi (p^2)\over m\beta (p^2)}={1\over 2Z_\pi G_V}.\cr}
\eqno(3.5)
$$
The comparison of the last two equations leads to
a matching condition which relates
physical quantities from different sectors of the model. In the
particular case we deal with,
properties of the $\rho$ meson and of the pion are related via
$$
\eqalign{
m^2_\rho =af^2_\rho f^2_\pi.\cr}
\eqno(3.6)
$$
The constant $a$ in our case is equal to
$$
\eqalign{
a={Z_\pi\over 2 m Z_\rho \beta_\pi }.\cr}
\eqno(3.7)
$$
For $a=2$ this is known as the KSFR relation [10].
The constant of the $\rho$ meson decay into two pions $f_{\rho\pi\pi}$
is usually included in the KSFR relation or eliminated from it as required
by the universality condition $f_{\rho\pi\pi}=f_\rho$. Let us see to
what extent the model behaviour of form factors agrees with this hypothesis.
For that, we calculate the form factor $f_{\rho\pi\pi}(p^2)$,
$$
f_{\rho\pi\pi}(p^2)={g_\rho (p^2)\over\sqrt{Z_{\rho}}}F(p^2).
\eqno(3.8)
$$
The function $F(p^2)$ represents the contribution of the
$\rho\rightarrow\pi\pi$ triangular diagrams (including \pa\ mixing effects)
and has the form
$$
\eqalignno{
F(p^2)={1\over Z_\pi}\biggl\{1-{m\beta_\pi p^2\over
       m^2_\rho (p^2)}+{1-2m\beta_\pi\over p^2-4m^2_\pi}
       \biggl[(p^2&-2m^2_\pi )\left({I_2(p^2)\over I_2(m^2_\pi )}-1
       \right)\cr
       &\quad +2m^4_\pi {I_3(-p_1, p_2)\over I_2(m^2_\pi )}
       \biggr]\biggr\} .&(3.9)\cr}
$$
Here $p=p_1+p_2$, with $p_1, p_2$ the pion momenta. From this one easily
deduces the following relation which holds  off the $\rho$ mass shell
$$
{f_{\rho\pi\pi}(p^2)\over f_\rho (p^2)}={F(p^2)\over Z_\rho}.
\eqno(3.10)
$$
In particular, at $p^2=0$ the equality
$$
I_2(0)-I_2(m^2_\pi )-m^2_\pi I_3(-p_1, p_2)\big\vert_{p^2=0}=
      2m^2_\pi {\partial I_2(p^2)\over \partial p^2}
      \bigg\vert_{p^2=m^2_{\pi}}
\eqno(3.11)
$$
leads to $F(0)=1$ and as a consequence one has
$f_{\rho\pi\pi}(0)=f_\rho (0)/Z_\rho$.

Let us consider now the properties of the axial-vector meson ($a_1$).
For this purpose we have to calculate the axial-vector form factor
$f_a(p^2)$ which describes the weak $a_1\rightarrow l\nu_l$ matrix element.
The relevant calculations give
$$
\eqalign{
{1\over f_a(p^2)}={2g_\rho (p^2)\over 3\sqrt{Z_a}}J_2(p^2)
                  [1-2m\beta (p^2)]=\sqrt{Z_\rho\over Z_a}
                  {1-2m\beta (p^2)\over f_\rho (p^2)},\cr}
\eqno(3.12)
$$
which leads to the result
$$
\eqalign{
{m^2_a(p^2)\over f^2_a(p^2)}={1-2m\beta (p^2)\over 4Z_a G_V},\cr}
\eqno(3.13)
$$
where we used
$$
\eqalign{
m^2_a(p^2)={m^2_\rho (p^2)\over 1-2m\beta (p^2)}.\cr}
\eqno(3.14)
$$
The relations (3.5) and (3.13) lead us to the another
matching condition
which relates the properties of the $a_1$ meson to pionic ones,
$$
\eqalign{
m^2_a=bf^2_af^2_\pi.\cr}
\eqno(3.15)
$$
The constant $b$ is equal to
$$
\eqalign{
b={Z_\pi (1-2m\beta_a)\over 2m Z_a \beta_\pi }.\cr}
\eqno(3.16)
$$
{}From the
relations (3.6) and (3.15) one can deduce a set of
equivalent formulae.
In particular, one of the direct consequences are the  well-known
Weinberg relations [11]. These are known to play a
crucial role in the calculation of the electromagnetic $\pi^+$--$\pi^0$
mass difference in the chiral limit. An analogue of Weinberg's first
sum rule is the equality
$$
\eqalign{
Z_a{m^2_a\over f^2_a}+Z_\pi {\beta_a\over \beta_\pi}f^2_\pi
  =Z_\rho {m^2_\rho\over f^2_\rho }.\cr}
\eqno(3.17)
$$
The second sum rule is described by the relation
$$
\eqalign{
Z_a{m^4_a\over f^2_a}-Z_\rho {m^4_\rho\over f^2_\rho }=
  {m^2_\rho (m^2_a)-m^2_\rho\over 4G_V}.\cr}
\eqno(3.18)
$$
The coefficients $Z, \beta$ and a non-zero right-hand side in (3.18)
result from determining physical properties of particles at different
values of $p^2$ corresponding to real values of their masses. In any
approach where all physical characteristics of the
mesons are fixed at a certain
scale these relations take the standard form (in this case $Z=1,
\beta_a=\beta_\pi , m^2_\rho (m^2_a)=m^2_\rho $). Off the mass shell
(3.18) is replaced by the equality
$$
\eqalign{
Z_a{m^4_a (p^2)\over f^2_a (p^2)}=
Z_\rho {m^4_\rho (p^2)\over f^2_\rho (p^2)}.\cr}
\eqno(3.19)
$$

\eject
\medskip
\noindent{\bf 4. Summary}
\medskip

Let us briefly summarize our results.
We have considered an extended NJL lagrangian with four quark interactions.
It includes some of the salient features of QCD in the low--energy
(long distance) regime. We have generalized the bosonization procedure
of Ref.[1] to the case with vector and axial-vector mesons. This allows
to calculate any mesonic N--point function to all orders in momenta.
In particular, we have
considered some important relations connecting pion properties
to ones of vector and axial-vector mesons. As specific examples
we have shown that a generalized
KSFR relation and generalized
Weinberg sum rules are obtained from matching conditions relating the
various sectors. These modified relations stem from a
consistent treatment of the
higher order terms in momenta. To lowest order
one recovers the celebrated results [10,11]. Finally,
            we note that similar issues were studied in ref. [12]. There,
however, all physical properties of mesons were fixed at $p^2=0$,
which leads to substantial differences in the final results.

\bigskip
\bigskip
\goodbreak
\centerline{\bf REFERENCES}
\bigskip
\item{[1]}  V.Bernard, A.A.Osipov and Ulf-G.Mei\ss ner, Phys.Lett.
            B285 (1992) 119.
\medskip
\item{[2]}  T.Eguchi, Phys.Rev. D14 (1976) 2755;

            K.Kikkawa, Progr.Theor.Phys. 56 (1976) 947.
\medskip
\item{[3]}  D.Ebert and M.K.Volkov, Yad.Fiz.36 (1982) 1265;
            Z.Phys. C16 (1983) 205.
\medskip
\item{[4]}  M.K.Volkov, Ann.Phys. 157 (1984) 282.
\medskip
\item{[5]}  A.Dhar and S.R.Wadia, Phys.Rev.Lett. 52 (1984) 959;

            A.Dhar, R.Shankar and S.R.Wadia, Phys.Rev. D31 (1985) 3256.
\medskip
\item{[6]}  D.Ebert and H.Reinhardt, Nucl.Phys. B271 (1986) 188;
            Phys.Lett. B173 (1986) 453.
\medskip
\item{[7]}  W.Pauli and F.Villars, Rev.Mod.Phys. 21 (1949) 434.
\medskip
\item{[8]}  V.Bernard and D.Vautherin, Phys.Rev. D40 (1989) 1615;

            C.Sch\"{u}ren, E.Ruiz Arriola and K.Goeke, Nucl.Phys.
            A547 (1992) 612.
\medskip
\item{[9]}  M.Gell-Mann, R.Oakes and B.Renner, Phys.Rev. 175 (1968) 2195.
\medskip
\item{[10]} K.Kawarabayashi and M.Suzuki, Phys.Rev.Lett. 16 (1966) 255;

            Riazuddin and Fayyazuddin, Phys.Rev. 147 (1966) 1071.
\medskip
\item{[11]} S. Weinberg, Phys.Rev.Lett. 18 (1967) 507.
\medskip
\item{[12]} J.Bijnens, E.de Rafael and H.Zheng, Preprint
            CERN-TH.6924/93 (1993).

\vfill
\eject
\nobreak
\end